\documentclass[%
 reprint,prb,
 amsmath,amssymb,
 aps,floatfix]{revtex4-2}

\usepackage{graphicx}
\usepackage{dcolumn}
\usepackage{bm}
\usepackage{hyperref}
\usepackage{multirow}
\usepackage{makecell}

\usepackage{ulem}
\usepackage{ xcolor }

\begin{document}

\title{Optical Properties and Spin States of Inter-layer Carbon Defect Pairs in Hexagonal Boron Nitride: A First-Principles Study}

\author{Ignacio Chac\'on}
\affiliation{Departamento de F\'isica, Facultad de Ciencias, Universidad de Chile. Santiago, Chile}
\author{Andrea Echeverri}
\affiliation{Departamento de F\'isica, Facultad de Ciencias, Universidad de Chile. Santiago, Chile}
\author{Carlos C\'ardenas}
\affiliation{Center for the Development of Nanoscience and Nanotechnology (CEDENNA), Santiago, Chile}
\affiliation{Departamento de F\'isica, Facultad de Ciencias, Universidad de Chile. Santiago, Chile}
\author{Francisco Munoz}
\affiliation{Center for the Development of Nanoscience and Nanotechnology (CEDENNA), Santiago, Chile}
\affiliation{Departamento de F\'isica, Facultad de Ciencias, Universidad de Chile. Santiago, Chile}
\email{fvmunoz@gmail.com}

\date{\today}

\begin{abstract}

Substitutional carbon defects in hexagonal boron nitride (hBN) are prominent single photon emitters (SPEs), and their potential for spin activity ($S\geq1$) is particularly intriguing. While studies have largely focused on intra-layer defects, we employ density functional theory (DFT) to investigate inter-layer dimers of identical carbon species (C$_X$C$_X$). We demonstrate that these C$_X$C$_X$ pairs can exhibit a stable triplet spin state at room temperature when closely spaced (\textit{e.g.}, within 3.5-7.1~\AA) across hBN layers. As their separation increases beyond this range (\textit{e.g.}, $>7$~\AA), they transition into weakly interacting $S=1/2$ pairs, characterized by singlet-triplet degeneracy. This regime is predicted to result in a very small zero-field splitting for the triplet manifold, offering a potential explanation for certain optically detected magnetic resonance (ODMR) signals. The zero-phonon line (ZPL) energy of these inter-layer C$_X$C$_X$ pairs is found to be practically monochromatic and within the visible range. Furthermore, we identify specific C$_B$C$_B$ inter-layer configurations exhibiting atypical low-energy phonon replicas due to out-of-plane vibrational coupling, a finding that may clarify the vibronic structure of other hBN emitters, such as the `yellow emitters'.
\end{abstract}

\maketitle

\section{\label{sec:intro}Introduction}
Hexagonal boron nitride (hBN) is a layered material that can be exfoliated down to the monolayer. It has garnered attention for hosting single photon emitters  \cite{Tran2015, Aharonovich2022Quantum, Qui2024}. Many of these defects have been associated with substitutional carbon defects \cite{Tran2015,mendelson2021, Weston2018, Linderalv2021, Mackoit2019, Du2015, Bourrellier2016, Wong2015, Golami2022, Fisher2023, Maciaszek2024Blue}, even though other options have been proposed \cite{Neumann2023organic, Cholsuk2023comprehensive, Kirchoff2022}. Due to its large bandgap, close to 6 eV, defects incorporated within the crystal can introduce new energy levels within the gap \cite{Cassabois2016, Rom_n_2021}. Consequently, electronic transitions give rise to SPEs exhibiting characteristics such as bright photons, a commendable coherence time \cite{Gottscholl2021}, even at room temperature, and in some cases, they are spin-active (triplet or higher spin state) defects \cite{Stern2022, Kianina20,li2022, Gao2021}. These properties are crucial for quantum technologies in communication, computation, metrology, sensing, imaging, etc. \cite{europe, Auburger21, Moon2023, Kianina2022, cakan2024quantum, Cholsuk2024Identifying, Esmann2024}. Therefore, the search for these types of defects is relevant, considering the scarcity of substitutional spin-active carbon defects, such as C$_N$C$_N$ \cite{Pinilla}, (C$_N$)$_3$C$_B$ and (C$_B$)$_3$C$_N$ \cite{Macia2022, Benedek2023}. It is worth mentioning that vacancy-based SPEs in hBN are an active research field, as they often are spin-active \cite{Guo2021, Liu2022, Qian2022Unveiling, Gottscholl2020}.

Substitutional C defects behave as donors (C$_B$), acceptors (C$_N$), donor-acceptor pairs, or more complex arrangements \cite{Huang2022, Almohammad2024Carbon}. Each C defect contributes a single electron and one defect state (spin up and down) within the hBN band gap. The 4.1 eV emission line has been identified as two adjacent C$_N$C$_B$ defects \cite{Mackoit2019}, often denoted as C$_2$. If the same C$_N$--C$_B$ pair is in the same layer but separated by some distance, the emission energy decreases to the visible range \cite{jara2021, Auburger21}. The ground state is likely spinless if the number of C$_N$ and C$_B$ is the same. Otherwise, the system can have a non-zero net spin. There are experimental findings of emissions attributed to C defects, even though the actual C defect has not been identified \cite{Stern2022, Scholten2024}. One possible explanation, fitting with the available results, is weakly coupled $S=1/2$ defects \cite{robertson2024universal}.

This article aims to investigate SPEs when forming a C dimer but residing in different hBN layers. Such a possibility has not been addressed before, even though defects comprising an interstitial C atom have been considered \cite{Bhang21}. The paper is structured as follows: Section~\ref{sec:methods} details the methods employed for the calculations, Section~\ref{sec:results} presents our main results, including the SPE's spin state, likely to be a triplet, and their photoluminescence spectra. Finally, Section~\ref{sec:discuss} explains the large energy difference between triplet and singlet states for some dimers, the phonon sidebands shape, and we relate our findings to experimental results on spin-active SPEs.

\section{\label{sec:methods}Methodology}
\subsection{\label{sec:computation} Computational Methods}

Density functional theory (DFT) calculations were performed using the Vienna ab initio simulation package (VASP) \cite{vasp1,vasp2,vasp3,vasp4}. We utilized a $7\times7$ hexagonal supercell of hBN with $\Gamma$-point sampling. Defects in adjacent layers were modeled using an hBN bilayer, while an hBN trilayer was employed for defects in non-adjacent layers. These multi-layer models aim to approximate bulk hBN; additional layers are expected to influence the zero-phonon line (ZPL) energy through \textit{(i)} enhanced dielectric screening \cite{amblard2022universal, winter2021, Badrtdinov2023Dielectric} and \textit{(ii)} modified electrostatic potentials \cite{Pinilla2024}, with these influences diminishing as layer thickness increases towards the bulk limit. The AA' stacking (B and N atoms on top of N and B, respectively), characteristic of most hBN samples \cite{Gilbert2019}, was adopted.

The kinetic energy cutoff was set to 500 eV. PyProcar \cite{pyprocar,pyprocar2} was used for the analysis of the results, and VESTA for visualization \cite{vesta}.

The SCAN exchange-correlation (XC) functional \cite{SCAN} was employed for structural relaxation, phonon calculations, and optical transition energies. The suitability of SCAN for defect levels is supported when the host's band gap is reasonably described \cite{Rauch2021}, as is the case here (SCAN-predicted hBN gap: 5.0 eV; experimental value $\approx 6$~eV). While the Heyd-Scuseria-Ernzerhof (HSE06) functional\cite{hse06} is a common alternative, its results are highly sensitive to dielectric screening, which is significantly reduced and anisotropic in 2D materials like hBN. This sensitivity is pronounced for donor-acceptor defects with strong electrostatic interactions. Consequently, applying HSE06 to our bilayer and trilayer models, which inherently possess different screening characteristics, yields results whose direct comparison or extrapolation to bulk behavior is complex. Nevertheless, HSE06 optical transition energies, calculated using the SCAN-relaxed geometries (a common practice where less demanding functionals provide reliable structures \cite{Deak2010}), are reported in the Supplementary Material. On the other hand, the functional of Perdew-Burke-Ernzerhof \cite{pbe} failed to describe the ground state of the defects at large distances, providing a spurious triplet state. Even worse, when using the PBE's wavefunctions as seeds for more expensive calculations, the error is propagated.  

To study the excited state of each configuration, the $\Delta$SCF method was used \cite{yang2024, Galli2021}. This method has a problem of correctly describing the excited state (\textit{i.e.} mixed spin states are employed instead of well-defined spins), and there are some corrections to improve the ZPL value \cite{Mackoit2019, iwanski2024revealing}. The most interesting systems in our study have a triplet ground state and involve transitions with the valence or conduction bands, so getting the correct Slater determinant of the excited state is too involved to be practical. In the Supplementary Material, we bound this error in the ZPL energy to $0.2$~eV or lower. In all cases, we restricted our study to neutral defects \cite{Macia2022}. Calculation of the single-photon emission rates would involve the computation of electric dipole transition moments between the involved excited and ground electronic states, which is beyond the scope of the current work.

The formation energy of a defect was calculated by using
\begin{equation}
\label{eq:eform}
\begin{split}
    H_f(D) =& E_{tot}(D) - E_{hBN} - n_C\mu_C \\ 
    &- n_B\mu_B - n_N\mu_N,
\end{split}
\end{equation}
where the $H_f(D)$ is the formation energy of a defect $D$, $E_{tot}(D)$ is the defect's energy, $E_{h-BN}$ is the energy of a pristine hBN bi- or tri-layer, $n_X$ is the number of a specific atom $X={C, B, N}$ which is added ($n>0$) or removed ($n<0$) from the defectless hBN system, and $\mu_X$ is the chemical potential of a $X$ atom.

The chemical potentials in Eq.~\eqref{eq:eform} depend on the experimental growth conditions of hBN. Bounds on the values of the chemical potentials of N and B can be established by ab inito thermodynamics method \cite{Scheffler2001}.  In this approach, it is assumed that the growth is a process in thermodynamic equilibrium and that the hBN serves as a reservoir of N and B. Therefore, the chemical potentials of B and N are not independent but linked by the formation enthalpy of hBN with respect to bulk B and gaseous \(N_2\) ($\mu_B+\mu_N=\Delta H_f(BN)=-2.6$ eV/formula unit) \cite{Tomaszkiewicz_2002,nist}. Hence, two bounds are in order. The N-poor condition in which $min(\mu_N)=\Delta H_f(BN)+1/2\mu_{N_2}^{gas}$, and the 
N-rich condition in which $min(\mu_B)=\Delta H_f(BN)+\mu_{B}^{bulk}$ \cite{Weston2018}. The chemical potential of carbon is taken as the C-rich bound such that $\mu_C^{graphite}$.

To calculate the phonon sideband of photoluminescence spectra, we used the method developed by Alkauskas \textit{et al.} \cite{alkauskas}. In summary, the phonons (assumed to be the same in the ground and excited states) are expanded into generalized coordinates $q_k$, resonating with the rearrangement upon emission.
\begin{equation}
    q_k = \sum_{\alpha}\sqrt{m_{\alpha i}}(R_{e,\alpha i} - R_{g,\alpha i})\Delta r_{k,\alpha i},
    \label{eq:g. coordinate}
\end{equation}
where $\alpha$, $i$, and $k$ denote atoms, Cartesian coordinates, and the vibrational mode, respectively. $R_e$ and $R_g$ are the positions in excited and relaxed states, respectively. $m_\alpha$ is the atom's mass. Finally, $\Delta r_k$ is the unitary vector for the $k$-th vibrational mode. Thus, we can obtain the partial Huang-Rhys factor $S_k$, representing the number of phonons of the phonon $k$: 
\begin{equation}
    S_k = \frac{\omega_k q_k^2}{2\hbar},
    \label{eq:Partial_HR}
\end{equation}
where $\omega_k$ is the frequency of the phonon $k$-th vibrational mode and $\hbar$ is the Planck constant. The total Huang-Rhys factor is $S=\sum_kS_k$. The photoluminescence spectral lineshape $L(h\nu)$ was then constructed using the generating function formalism, which builds the vibronic progression based on $E(ZPL)$, the set of $\{S_k\}$, and $\{\hbar\omega_k\}$ \cite{alkauskas}. Finally, the calculated stick spectrum (ZPL and phonon replicas) was convoluted with a Gaussian function with a Full Width at Half Maximum (FWHM) of 25 meV. This phenomenological broadening accounts for homogeneous lifetime effects, inhomogeneous broadening, and approximates thermal effects at room temperature, allowing for comparison with experimental spectra.

From here, the optical spectral function and the phonon sideband can be directly determined \cite{alkauskas}. The only free parameter is a thermal-like broadening of $S_k$, which we set to 25 meV (room temperature).

\section{\label{sec:results}Results}

%\subsection{\label{sec:config} Defect Geometry}

\begin{figure}[ht]
    \centering
    \includegraphics[width=\columnwidth]{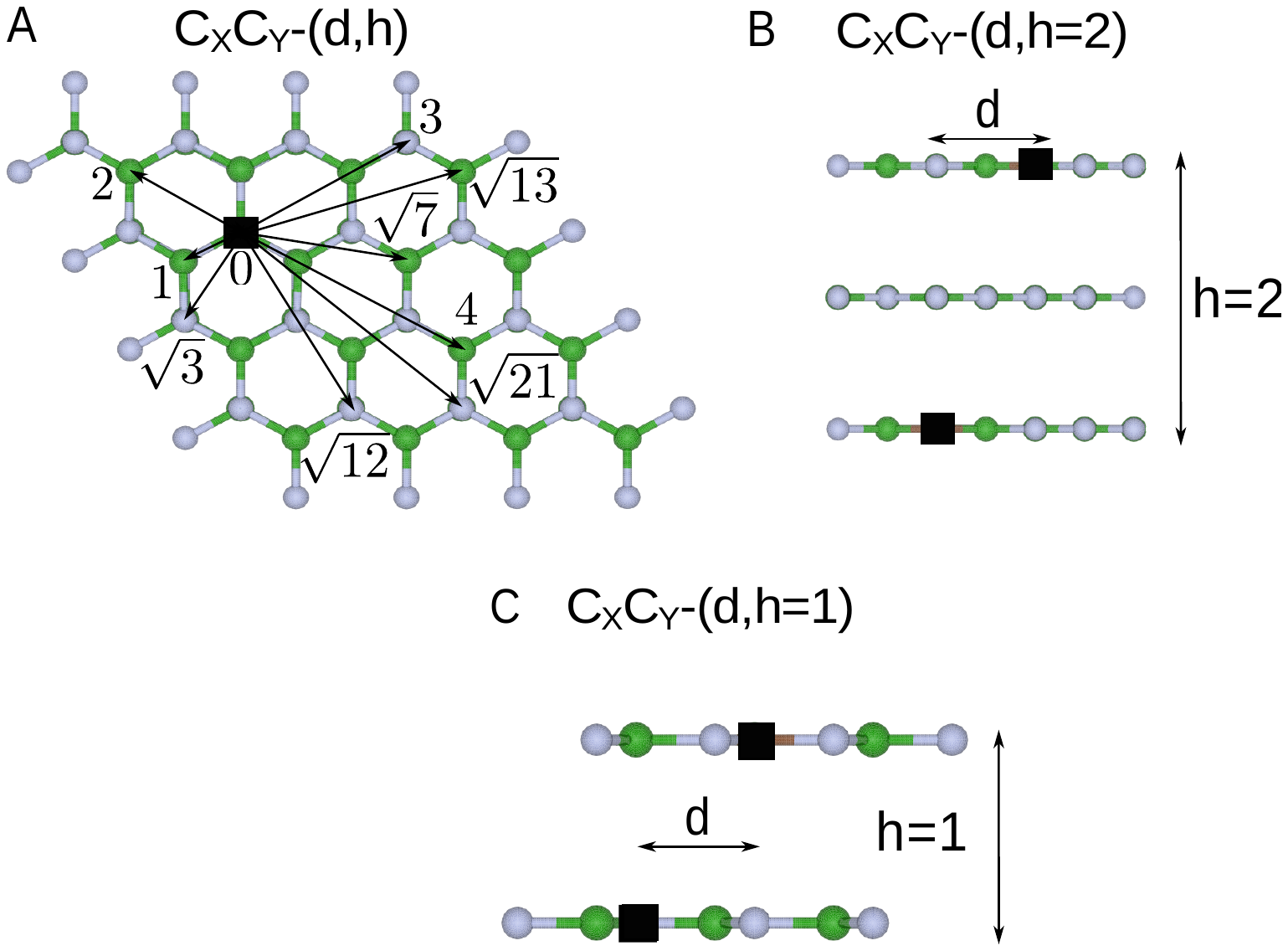}
    \caption{Geometry of the studied defects, neutral dimers C$_X$C$_Y$ with $X,Y = \{$N, B$\}$. The defects are separated by an in-plane distance $d$ plus an out-of-plane distance $h$. Panel A shows $d$ for the defects studied, $d$ is measured in times the nearest neighbor distance. In the figure one C atom is at the black square, and the other is at the end of the arrow. Table~\ref{tab:distances} gives the in-plane distance in \AA~ for each value of $d$. Panels B and C show the out-of-plane distance $h$, with $h=1$ if the C atoms (black squares) are in adjacent layers and $h=2$ when there is an hBN among the defect atoms. The different colors of the hBN lattice represent B and N atoms without specifying them (\textit{i.e.} if the C atom in panel A is $C_B$, N is green; if the C atom is $C_N$, B is green).}
    \label{fig:configuration}
\end{figure}

All of the defects studied are substitutional C-dimers, with both C atoms in different layers. The cases of C$_N$C$_B$, C$_N$C$_N$ and C$_B$C$_B$ were studied. In all dimers, the C atom in the bottom layer was kept fixed while we changed the position of the C atom in the top layer, see Fig.~\ref{fig:configuration}A. For clarity, we labeled the in-plane position ($d$) of one C atom with respect to the other by the in-plane projection of their distance, taking the nearest neighbor distance as `1' (\textit{i.e.} $d=1$ is equivalent to 1.45 \AA), as found in a similar context for labeling C-defects \cite{Auburger21}. To convert $d$ to times the lattice parameter, it is needed to multiply it by $\sqrt{3}$. The conversion of $d$ to \AA{} is given in Table~\ref{tab:distances}. The C atoms can belong to adjacent layers ($h=1$) or can be separated by an inner layer ($h=2$); see Fig.~\ref{fig:configuration}B-C. Thus, a given dimer defect can be identified by C$_X$C$_Y$--$(d,h)$, with $X,Y=\{$B, N$\}$.

\begin{table}[ht]
\begin{tabular}{|c|c|cc|}
\hline
\multirow{2}{*}{\makecell{Distance $d$ \\ (nearest neighbors)}} & \multirow{2}{*}{Distance $d$ (\AA)} & \multicolumn{2}{c|}{Distance C-C (\AA)} \\ \cline{3-4}  &  & \multicolumn{1}{c|}{$h=1$}  & $h=2$  \\ \hline
$0$          & $0$    & \multicolumn{1}{c|}{$3.28$}   & $6.68$   \\ \hline
$1$          & $1.45$ & \multicolumn{1}{c|}{$3.58$}   & $6.85$   \\ \hline
$\sqrt{3}$   & $2.50$ & \multicolumn{1}{c|}{$4.12$}   & $7.14$   \\ \hline
$2$          & $2.90$ & \multicolumn{1}{c|}{$4.38$}   & $7.28$   \\ \hline
$\sqrt{7}$   & $3.82$ & \multicolumn{1}{c|}{$5.04$}   & $7.71$   \\ \hline
$3$          & $4.31$ & \multicolumn{1}{c|}{$5.43$}   & $7.96$   \\ \hline
$\sqrt{12}$  & $5.03$ & \multicolumn{1}{c|}{$6.03$}   & $8.36$   \\ \hline
$\sqrt{13}$  & $5.23$ & \multicolumn{1}{c|}{$6.19$}   & $8.49$   \\ \hline
$4$          & $5.80$ & \multicolumn{1}{c|}{$6.67$}   & $8.85$   \\ \hline
$\sqrt{21}$  & $6.62$ & \multicolumn{1}{c|}{$7.39$}   & $9.40$   \\ \hline
\end{tabular}
\caption{The in-plane component of the distance, $d$, and the total distance for each pair of defects C$_X$C$_Y$--$(d,h)$. The labels are defined in Fig.~\ref{fig:configuration}.}
\label{tab:distances}
\end{table}

%\subsection{\label{sec:spin} Spin State}

\begin{figure}[ht]
    \centering
    \includegraphics[width=\columnwidth]{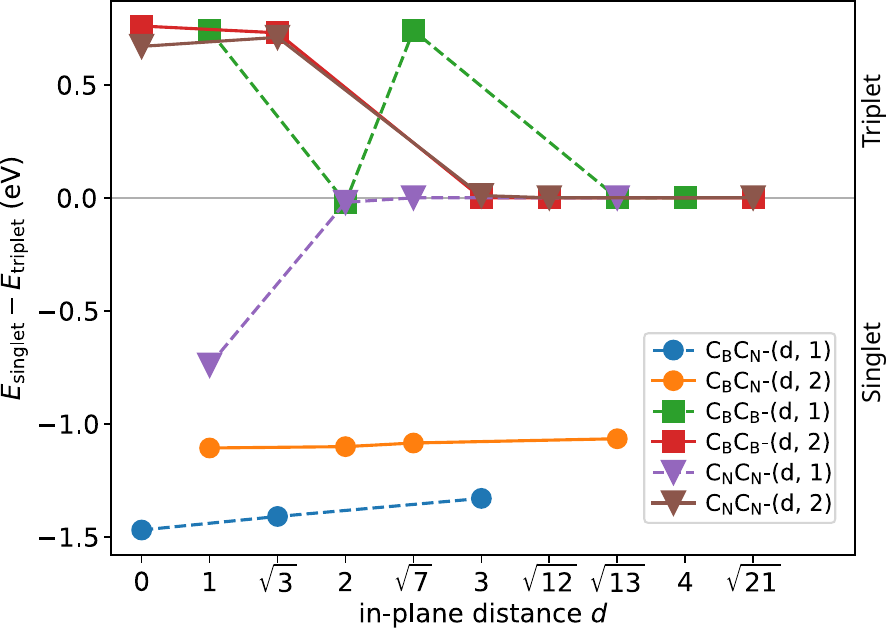}
    \caption{Energy difference between the singlet and triplet state of each configuration. A positive (negative) value means a favorable triplet (singlet) state. }
    \label{fig:ST}
\end{figure}

Defects of the type C$_N$C$_B$ are consistently found in our calculations to exhibit a singlet ground state (see Fig.~\ref{fig:ST}). This observation aligns with established expectations: C$_N$ sites, characterized by defect energy levels close to the valence band maximum, act as acceptors. Conversely, C$_B$  sites are understood to behave as donors. Consequently, a charge transfer from C$_B$ to C$_N$ is anticipated, irrespective of their separation, leading to effectively charged C$_B^{\delta+}$ and C$_N^{\delta-}$ centers. In the dissociation limit, as direct quantum mechanical interactions (such as exchange coupling) between the two carbon atoms vanish, their distinct electronic natures --C$_B$ as a donor and C$_N$ as an acceptor-- remain the dominant factor. Overall, the system will naturally remain a singlet, consistent with our findings.

In contrast, carbon defects of the same species, \textit{i.e.}, C$_N$C$_N$ and C$_B$C$_B$, generally exhibit a stable triplet ground state when the defects are relatively close. This tendency is typically observed for separations up to $d=\sqrt{3}$ (an in-plane distance of $\sim 2.51$~\AA) and $h=2$ (a vertical separation of two layers, resulting in a total distance of approximately 7\AA), as detailed in Fig.~\ref{fig:ST}. However, for C$_B$C$_B$ defects in adjacent layers ($h=1$), the spin state at shorter distances can be more sensitive to the specific geometric configuration.

The spin state of the defect pair is intrinsically linked to the symmetry of the two-electron spatial wavefunction: a singlet state is associated with a spatially symmetric wavefunction, while a triplet state corresponds to a spatially anti-symmetric wavefunction. This spatial symmetry dictates the charge distribution of the pair; for instance, an anti-symmetric wavefunction (triplet) possesses a nodal plane or region of zero electron density at the midpoint between the defects, whereas a symmetric wavefunction (singlet) typically has finite electron density there. This difference in charge distribution influences the electrostatic interaction with the polar hBN lattice, a mechanism previously invoked to explain the pronounced stability of the triplet state for certain C$_N$C$_N$ configurations within the same hBN layer (h=0) \cite{Pinilla}.

As the defects are moved further apart, the significance of the wavefunction's symmetry at the midpoint diminishes because the overall wavefunction amplitude in this central region approaches zero. In this dissociation limit, the C atoms evolve into weakly interacting S=1/2 paramagnetic centers \cite{robertson2024universal}. Consequently, the exchange interaction between them vanishes, leading to the singlet and triplet states becoming effectively degenerate, as anticipated for two distant, uncorrelated S=1/2 spins.

A notable exception to the general triplet preference for nearby C$_N$C$_N$ defects is the C$_N$C$_N$--(1,1) configuration. In this case, the two C$_N$ atoms experience a strong mutual attraction, causing them to displace significantly from the hBN plane and form a direct covalent C-C bond. This bonding scenario, which locally resembles sp$^3$ hybridization, results in a singlet ground state, \textit{i.e.} consistent with the spin-pairing typically observed upon formation of a strong covalent bond.

\begin{table}[ht]
\caption{Formation energy $H_f$ (in eV) of the various C$_X$C$_Y$ defect configurations studied. Energies are reported for two extreme chemical potential environments: N-poor (B-rich) and N-rich. Defects are grouped by type (C$_B$C$_B$, C$_N$C$_N$, C$_N$C$_B$) and then by vertical separation h (in layers) to guide comparison.}
\label{tab:F.E}
\begin{tabular}{|l|cc|}
\hline
\multirow{2}{*}{Configuration} & \multicolumn{2}{c|}{H$_f$ (eV)}       \\ \cline{2-3} 
                               & \multicolumn{1}{c|}{N-poor} & N-rich \\ \hline
C$_B$C$_B$--$(1,1)$            & \multicolumn{1}{c|} {9.05}       & 3.85
\\ \hline
C$_B$C$_B$--$(\sqrt{7},1)$     & \multicolumn{1}{c|} {9.08}       & 3.88       \\ \hline
C$_B$C$_B$--$(2,1)$            & \multicolumn{1}{c|} {9.08}       & 3.88       \\ \hline
C$_B$C$_B$--$(\sqrt{13},1)$    & \multicolumn{1}{c|} {9.08}       & 3.88       \\ \hline
C$_B$C$_B$--$(4,1)$            & \multicolumn{1}{c|} {9.07}       & 3.87       \\ \hline\hline
C$_B$C$_B$--$(0,2)$            & \multicolumn{1}{c|} {9.07}       & 3.87       \\ \hline
C$_B$C$_B$--$(\sqrt{3},2)$     & \multicolumn{1}{c|} {9.06}       & 3.86       \\ \hline
C$_B$C$_B$--$(3,2)$            & \multicolumn{1}{c|} {9.06}       & 3.86       \\ \hline
C$_B$C$_B$--$(\sqrt{12},2)$    & \multicolumn{1}{c|} {9.05}       & 3.85       \\ \hline
C$_B$C$_B$--$(\sqrt{21},2)$    & \multicolumn{1}{c|} {9.06}       & 3.86       \\ \hline\hline
C$_N$C$_N$--$(1,1)$            & \multicolumn{1}{c|} {2.88}       & 8.08       \\ \hline
C$_N$C$_N$--$(2,1)$            & \multicolumn{1}{c|} {3.47}       & 8.67       \\ \hline
C$_N$C$_N$--$(\sqrt{7},1)$     & \multicolumn{1}{c|} {3.44}       & 8.64       \\ \hline
C$_N$C$_N$--$(\sqrt{13},1)$    & \multicolumn{1}{c|} {3.45}       & 8.65       \\ \hline\hline
C$_N$C$_N$--$(0,2)$            & \multicolumn{1}{c|} {3.45}       & 8.65      \\ \hline
C$_N$C$_N$--$(\sqrt{3},2)$     & \multicolumn{1}{c|} {3.42}       & 8.62       \\ \hline
C$_N$C$_N$--$(3,2)$            & \multicolumn{1}{c|} {3.40}       & 8.60       \\ \hline
C$_N$C$_N$--$(\sqrt{12},2)$    & \multicolumn{1}{c|} {3.40}       & 8.60       \\ \hline
C$_N$C$_N$--$(\sqrt{21},2)$    & \multicolumn{1}{c|} {3.41}       & 8.61       \\ \hline\hline
C$_N$C$_B$--$(0,1)$            & \multicolumn{1}{c|} {4.71}       & 4.71       \\ \hline
C$_N$C$_B$--$(\sqrt{3},1)$     & \multicolumn{1}{c|} {4.75}       & 4.75       \\ \hline
C$_N$C$_B$--$(3,1)$            & \multicolumn{1}{c|} {4.84}       & 4.84       \\ \hline\hline
C$_N$C$_B$--$(1,2)$            & \multicolumn{1}{c|} {5.13}       & 5.13       \\ \hline
C$_N$C$_B$--$(2,2)$            & \multicolumn{1}{c|} {5.14}       & 5.14       \\ \hline
C$_N$C$_B$--$(\sqrt{7},2)$     & \multicolumn{1}{c|} {5.15}       & 5.15       \\ \hline
C$_N$C$_B$--$(\sqrt{13},2)$    & \multicolumn{1}{c|} {5.17}       & 5.17       \\ \hline
\end{tabular}
\end{table}

The formation energies ($H_f$) of the investigated C$_X$C$_Y$ defects are presented in Table~\ref{tab:F.E}. For C$_B$C$_B$ and C$_N$C$_N$ defects, $H_f$ is primarily determined by the chemical environment (\textit{i.e.}, the chemical potential of nitrogen). Under their most favorable conditions --N-rich for C$_B$C$_B$ and N-poor for C$_N$C$_N$-- the formation energies for C$_B$C$_B$ configurations are found to be in the narrow range of 3.85-3.88 eV, while for C$_N$C$_N$ configurations they range from 2.88 eV to 3.47 eV. Conversely, in their least favorable environments --N-poor for C$_B$C$_B$ and N-rich for C$_N$C$_N$-- $H_f$ increases substantially, to approximately 9.05-9.08 eV for C$_B$C$_B$ and 8.08-8.67 eV for C$_N$C$_N$.

Generally, for these homonuclear pairs (C$_B$C$_B$ and C$_N$C$_N$), variations in the specific C--C distance ($d$) or interlayer arrangement ($h=\{1,2\}$) have a relatively minor impact on $H_f$ (typically less than 0.05 eV), especially when compared to the large shifts induced by the chemical environment. A prominent exception to this trend is the C$_N$C$_N$--(1,1) configuration. It exhibits the lowest formation energy 
among all C$_N$C$_N$ defects under both N-poor (2.88 eV) and N-rich (8.08 eV) conditions. This enhanced stability, representing a reduction of approximately 0.5-0.6 eV compared to other C$_N$C$_N$ configurations at $h=1$, is consistent with the formation of a stabilizing covalent C--C bond, as discussed in relation to its structure and spin state.

For the heteronuclear C$_N$C$_B$ defects, the calculated formation energies (Table~\ref{tab:F.E}) range from approximately 4.71 eV to 5.17 eV. These formation energies are identical under N-poor and N-rich conditions because the creation of a C$_N$C$_B$ pair involves substituting one B and one N atom with two C atoms; thus, the terms related to $\mu_N$ and $\mu_B$ in the formation energy expression effectively cancel each other. The formation energies for the inter-layer C$_N$C$_B$ configurations investigated here (all with $h=1$ or $h=2$, and not forming direct covalent C-C bonds between the pair) are substantially larger than the $H_f\approx 2.1$ eV reported for an adjacent C$_N$-C$_B$ pair within the same hBN layer ($h=0$) \cite{Macia2022}. The significantly lower energy of this intra-layer C$_2$ dimer is likely attributable to the formation of a direct C--C covalent bond, a strong stabilizing effect similar to that observed for the C$_N$C$_N$--(1,1) configuration in our study.

For broader context, other types of carbon defects, such as topological C$_N$C$_B$ defects (\textit{e.g.}, Stone-Wales-like structures), can exhibit even higher formation energies than the substitutional C$_N$C$_B$ pairs in Table~\ref{tab:F.E}, although their $H_f$ can be significantly reduced by mechanical strain \cite{babar2024carbon}. Indeed, strain engineering through biaxial stress has been shown to modulate the formation energies of individual C$_B$ or C$_N$ defects \cite{Santra2024}. While large, uniform strains are not typically expected in bulk hBN, localized stresses, for instance near grain boundaries, could influence defect formation and stability, with carbon defects potentially acting to relieve such stresses \cite{babar2024carbon}.

The overall thermal stability and persistence of these C defects in hBN are governed not only by their formation energies but also critically by the kinetic barriers against their migration or transformation. For instance, Babar \textit{et al.} \cite{babar2024carbon} calculated a high energy barrier (over 6 eV) for the transformation between a Stone-Wales C$_2$ dimer and a standard substitutional C$_2$ dimer, highlighting that C-C bond rearrangements can be significantly hindered. While specific migration and transformation pathways for the C$_X$C$_Y$--($d,h$) configurations in our study have not been computed, similar considerations of strong covalent bond stability suggest that their diffusion and spontaneous restructuring are also likely to be impeded by substantial energy barriers, limiting thermally induced migration under typical experimental conditions.

To further underscore the expected kinetic stability of these C--C pairs, we can estimate the energy cost associated with a specific decomposition mechanism. Consider a C$_B$C$_B$ pair, for instance, C$_B$C$_B$--($d,h=1$). One possible route for its partial decomposition involves one of the carbon atoms transitioning from its boron substitutional site (leaving it as V$_B$) into an interstitial position (C$_i$), while the other carbon atom remains as a C$_B$. The products of such a process would be an isolated C$_B$ defect, a carbon interstitial C$_i$, and a boron vacancy V$_B$. We can estimate the formation energy of this dissociated state, $H_f(\mathrm{C}_B) + H_f(\mathrm{C}_i) + H_f(\mathrm{V}_B)$, using the formation energies of these individual neutral defects. 
\begin{itemize}
    \item Under N-poor conditions, with $H_f(\mathrm{C}_B)\approx 4.5$ eV \cite{Bhang21}, $H_f(\mathrm{C}_i)=7.5$ eV \cite{Bhang21}, and $H_f(\mathrm{V}_B)=9.0$ eV \cite{Weston2018}, the total formation energy of the dissociated state is $\approx 21.0$ eV.
    \item Under N-rich conditions, with $H_f(\mathrm{C}_B)=1.8$ eV \cite{Bhang21}, $H_f(\mathrm{C}_i)=7.5$ eV \cite{Bhang21}, and $H_f(\mathrm{V}_B)=8.4$ eV \cite{Weston2018}, the total formation energy of this state is $\approx 17.7$ eV.
\end{itemize}
These values render the proposed decomposition pathway highly improbable, even at high temperature (\textit{i.e.}, a barrier over 10 eV). While $H_f(\mathrm{C}_B)$ is about one half of the formation energy of a C$_B$ dimer, the interstitial and the vacancy have a much higher $H_f$, which is expected, as in both defects, strong covalent $\sigma$ bonds become dangling bonds. Beyond such decomposition mechanisms, local structural transformations could also occur. For example, an in-plane rotation of a C$_B$-C$_N$ bond pair, a process analogous to a step in Stone-Wales defect formation, would constitute another pathway for atomic rearrangement. Again, covalent bonds need to be broken and reformed, and a very high barrier is expected for such processes in hBN \cite{babar2024carbon}.

A similar rationale would apply to the decomposition of C$_N$C$_N$ pairs, provided the formation energies of the analogous dissociated products (\textit{e.g.}, C$_N$, C$_i$, V$_N$) result in a similarly high overall energy cost. This analysis, while specific to one decomposition channel, supports the general expectation that transforming these substitutional carbon pairs into defect configurations involving interstitials and vacancies is kinetically strongly hindered. Overall, the robustness of the covalent network in hBN implies that most pathways for significant structural alteration or migration of these carbon defect pairs will involve high activation energies.

\begin{figure}[ht]
    \centering
    \includegraphics[width=0.95\columnwidth]{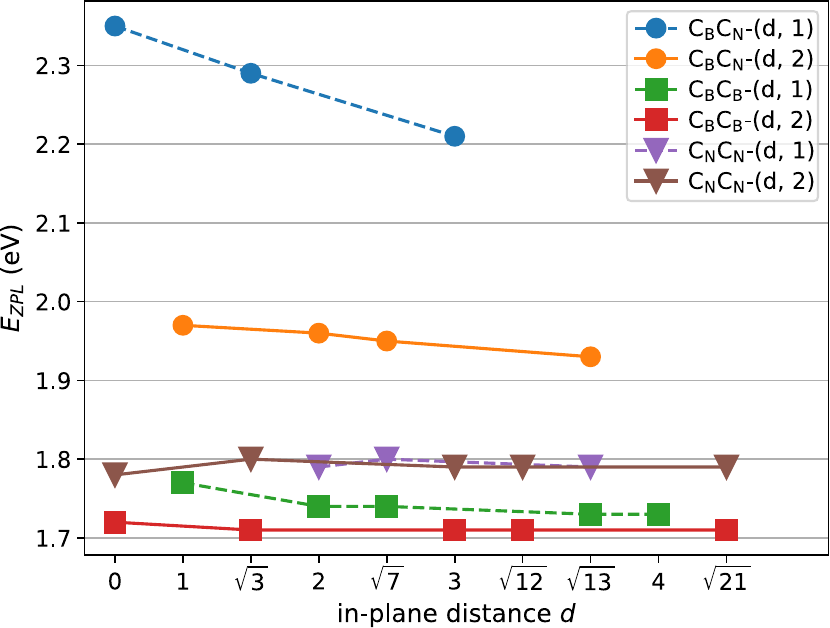}
    \caption{ZPL Energy of the configurations, calculated with the SCAN functional. There is no emission for C$_N$C$_N$--$(1,1)$, since there is a covalent C--C bond in this defect, precluding it from being a SPE. See  Fig.~S1,in the supplementary material for HSE values.}
    \label{fig:ZPL}
\end{figure}

The ZPL energy for the defects studied is shown in Fig.~\ref{fig:ZPL}, as calculated with the SCAN XC functional. In the supplementary material, Fig.~S1 shows the values calculated with HSE06 (see the Computational Methods). The donor-acceptor-like dimers have a ZPL energy decreasing with the C--C distance. As the individual defects are charged, their electrostatic interaction is considerable even at long distances. The emission should be in the green-orange range of the visible spectrum. The use of HSE06 is inconsistent for these heteronuclear defects, as we used a different thickness for $h=1,h=2$, effectively changing the dielectric constant. 

Defect dimers of the same type, C$_N$C$_N$ and C$_B$C$_B$, have a ZPL practically independent of the C--C distance (the smallest distance considered is over 3.5~\AA). For C$_B$C$_B$ defects, the ZPL is close to $1.72$ eV, and for C$_N$C$_N$ it is close to $1.80$ eV, see Fig.~\ref{fig:ZPL}. To have a reference value, with the HSE06 hybrid functional (Fig.~1 in the supplementary information), the ZPL energies of C$_B$C$_B$ and C$_N$C$_N$ are 2.1 and 2.4 eV, respectively. Again, the values are almost independent of the C--C distance. It is worth mentioning that, as there is no strong electrostatic interaction, this time the values of the HSE06 functional are not heavily affected by changing the dielectric constant. It is well documented that DFT has an inherent problem in determining accurately the band gap of materials \cite{Perdew1985}. This is particularly true for defect states \cite{Reimers2018}. Then, additional information, such as the vibrational spectra, is useful to determine the defect nature against experimental data \cite{jara2021}.

The C$_N$C$_N$--(1,1) configuration, which is characterized by the formation of a direct C-C covalent bond, was considered a distinct case. Its optical electronic transition involving this strong C-C bond (\textit{e.g.}, $\sigma\to\sigma^*$) should occur at energies larger than the band gap of hBN. For this reason, its ZPL associated with such deep defect states was not computed.

%%%% ACA VOY

\begin{figure}[ht]
    \centering
    \includegraphics[width=\columnwidth]{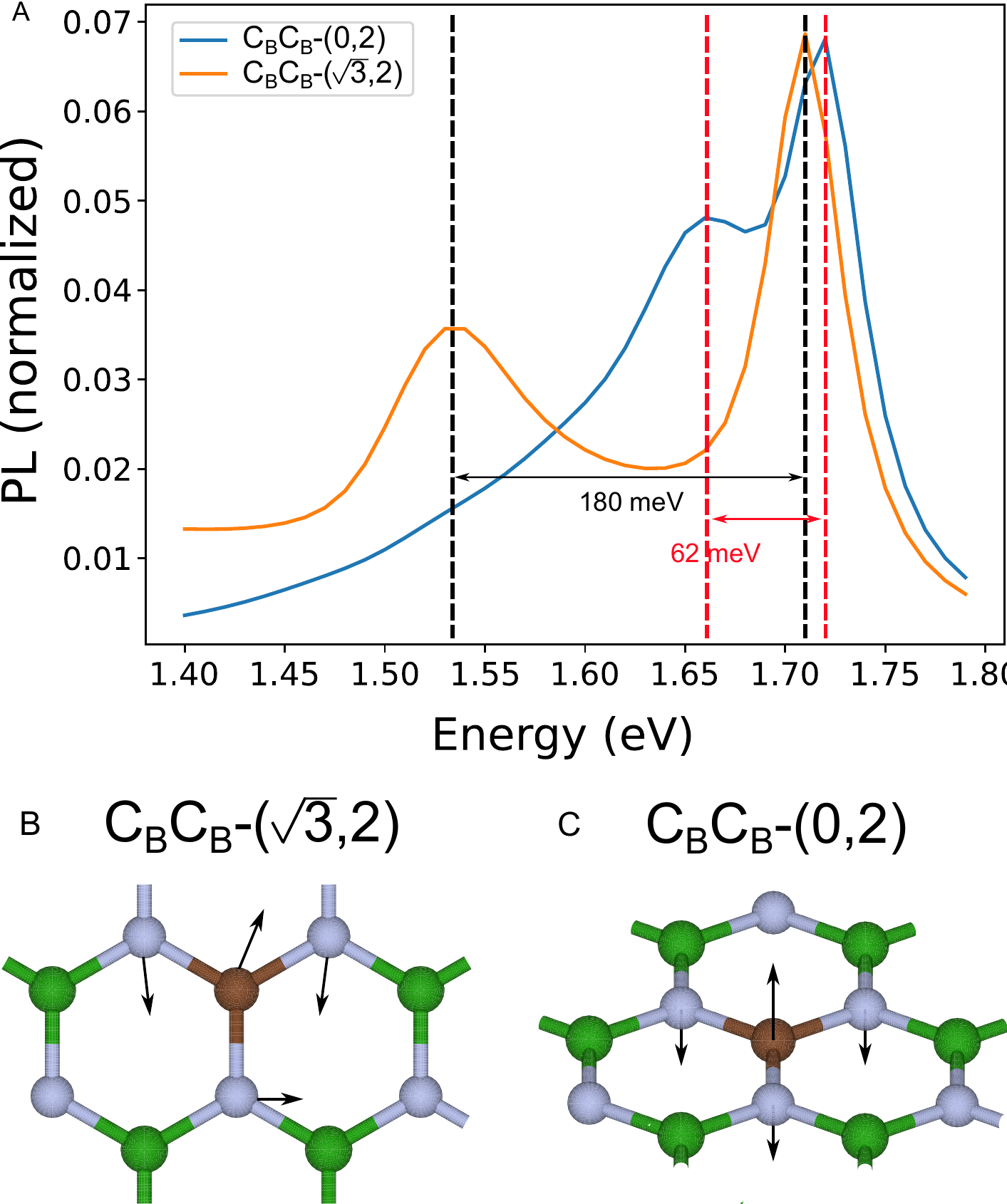}
    \caption{(A) Photoluminescence spectra of C$_B$C$_B$--$(0,2)$ (blue line) and C$_B$C$_B$--$(\sqrt{3},2)$ (orange line) defects. The first phonon replica is highlighted as a vertical dashed line. It is 62~meV for C$_B$C$_B$--$(0,2)$ and 180~meV for C$_B$C$_B$--$(\sqrt{3},2)$. (B,C) Phonons with the largest partial Huang-Rhys factor, $S_k$. In panel B (C), the displacements are in-plane (out-of-plane). Only one layer is in the figures, as both are connected by inversion symmetry. In panel B (C), the $C_3$ symmetry is broken (preserved) by the C$_B$ defects.}
    \label{fig:Phonon}
\end{figure}

Two types of photoluminescence spectra were found, see Fig.~\ref{fig:Phonon}A. The one associated with the defect C$_B$C$_B$--$(\sqrt{3},2)$ represents most of the defects studied here. Figures S2--S4 in the supplementary material show the photoluminescence spectra of all the studied defects emitting within the visible range. These high-energy phonon replicas are typically found in C-based defects in hBN \cite{jara2021,Tran2015}. They correspond to (in-plane) bond stretching involving the C atoms and their closest neighbors, see Fig.~\ref{fig:Phonon}B. While so far, only defects and defect clusters within a single layer had been considered, the mechanism behind the activation of this phonon mode should apply regardless of the exact location of both C atoms: one electron jumps between defect atoms, or from/to a defect level and the host bands, strengthening or weakening some bonds and causing the in-plane stretching. 

However, there are a couple of exceptions, C$_B$C$_B$-- $(0,2)$ (shown in Fig.~\ref{fig:Phonon}C), and  C$_B$C$_B$--$(1,1)$ with prominent phonon replicas of $\sim 62$ meV. In these cases, an out-of-plane phonon mode resonates with the transition. In the next section, we will discuss the origin of both types of phonon sidebands found.

\section{\label{sec:discuss}Discussion}
\subsection{\label{sec:Triplet}Triplet or singlet spin ground state}

\begin{figure}[ht]
    \centering
    \includegraphics[width=\columnwidth]{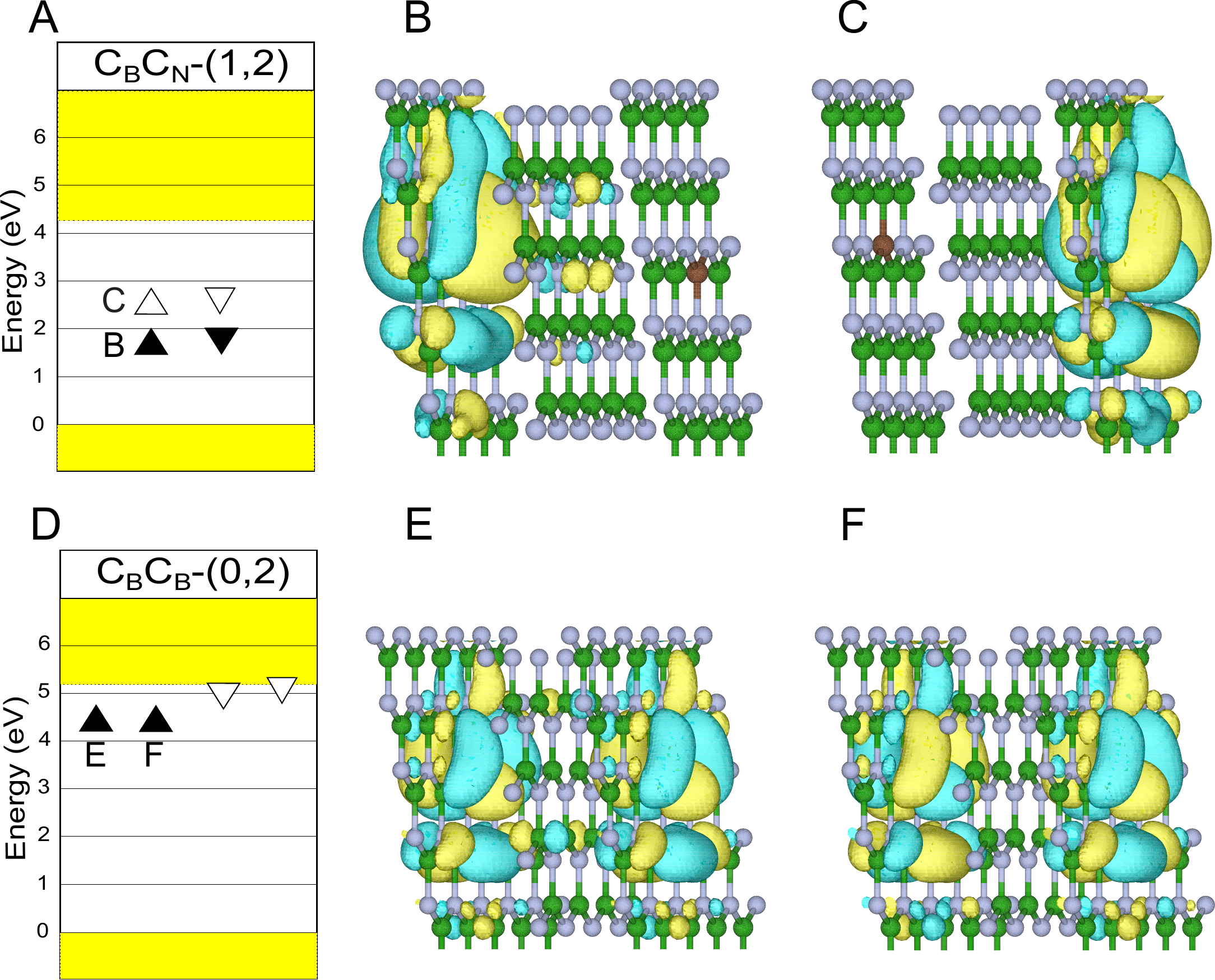}
    \caption{In panels A and D, the energy levels of the defects C$_B$C$_N$--$(1,2)$ in its singlet state and C$_B$C$_B$-$(0,2)$ in its triplet state are shown, respectively. The letters label specific levels, depicted in panels B, C, E, and F. Panels B and C illustrate the wavefunctions of the occupied C$_N$ and empty C$_B$ energy levels, respectively. Panels E and F display the wavefunctions of the C$_B$ occupied defect levels.
    }
    \label{fig:wavefunction}
\end{figure}

The relation between the spin state and the type of defect pair can be understood by examining the energy and nature of the defect levels within the fundamental band gap of hBN. As established, C$_B$ sites introduce donor-like defect states, positioned closer to the conduction band minimum. In contrast, C$_N$ sites create acceptor-like defect states, situated nearer to the valence band maximum. This distinction is illustrated schematically in panels A (for a C$_N$C$_B$ type) and D (for a C$_B$C$_B$ type) of Fig.~\ref{fig:wavefunction}, using the C$_B$C$_N$--$(1,2)$ and C$_B$C$_B$--$(0,2)$ configurations as respective examples.

Detailed electronic band structures for many of the defect configurations studied are provided in Figures {S5-S7} of the Supplementary Material. For cases where singlet and triplet states of a defect pair were found to be nearly degenerate in energy, only the band structure for the triplet state is presented, as the singlet counterpart yields almost identical results. Generally, the donor/acceptor picture for C$_B$/C$_N$ holds as long as the carbon defects maintain their sp$^2$ hybridization within the hBN plane. A notable exception is the C$_N$C$_N$--$(1,1)$ configuration (see Fig. S7 in the Supplementary Material), which undergoes significant structural rearrangement to form a C-C bond with sp$^3$-like character. For this defect, a single, non-degenerate defect-related band appears within the hBN band gap, with other associated levels located deep within the valence band.

For heteronuclear C$_N$C$_B$ defect pairs, the inherent donor (C$_B$) and acceptor (C$_N$) nature of the constituent defects leads to charge transfer from C$_B$ to C$_N$, resulting in C$_B^+$ and C$_N^-$ ionic states. This is depicted in Fig.~\ref{fig:wavefunction}A, where the occupied defect levels (associated with C$_N^-$) are shown lower in the gap, and empty levels (associated with C$_B^+$) are higher. The corresponding wavefunctions (Fig.~\ref{fig:wavefunction}B,C) illustrate the spatial localization of these occupied and empty states primarily on the C$_N$ and C$_B$ sites, respectively. This charge transfer results in a doubly occupied acceptor state localized on the C$_N$ site (forming C$_N^-$) and an emptied donor site (forming C$_B^+$). As these resulting ionic centers (C$_N^-$ and C$_B^+$) are effectively non-magnetic due to their spin-paired electronic configurations, the C$_N$C$_B$ defect pair consequently exhibits a singlet ground state.

In contrast, for homonuclear defect pairs, such as C$_B$C$_B$ (or C$_N$C$_N$), both components are of the same type (\textit{e.g.}, both donors for C$_B$C$_B$). For these homonuclear pairs, the identical nature of the two defects means their respective unperturbed defect states are intrinsically degenerate. When these defects interact at sufficiently short distances, this degeneracy is lifted, and new combined states are formed which can be described as linear combinations of the original defect states, as illustrated for C$_B$C$_B$ in Fig.~\ref{fig:wavefunction}D-F. For these systems, each C atom contributes an electron to the defect-related manifold. The significant overlap between the wavefunctions of these occupied levels gives rise to an exchange interaction. This interaction typically stabilizes the triplet state (parallel spins), where the spatial wavefunction is antisymmetric, relative to the singlet state (antiparallel spins), which has a symmetric spatial wavefunction, thus making the triplet state the ground state at shorter distances. As the separation between the identical defects increases, the wavefunction overlap and the consequent exchange interaction diminish. In this limit, the energy difference between the singlet and triplet states vanishes, leading to their degeneracy as the system effectively becomes two weakly interacting $S=1/2$ centers.

\subsection{Phonon sideband}

When examining the electron-phonon coupling of the optical transition, the relevant phonon replicas are high-energy modes corresponding to bond stretching, typically associated with substitutional C defects \cite{jara2021}. Fig.~\ref{fig:Phonon}A shows the defect C$_B$C$_B$--$(\sqrt{3},2)$, whose phonon sideband is representative of most of the defects studied here, {Figures S2-S4} in the supplementary material have the photoluminescence spectra for all remaining configurations. The optical excitation induces the transfer of one electron from a C atom to the other if the defect is C$_N$C$_B$, or from/to the C atom to the valence/conduction band if the defect is C$_X$C$_X$. Either way, the C--B and/or C--N bonds become weaker or stronger, resulting in a bond stretching: a high-energy localized longitudinal phonon. 

However, there are two defects with a markedly different photoluminescence spectra, C$_B$C$_B$--$(0,2)$ and  C$_B$C$_B$--$(1,1)$. In the former, the C$_B$ defects are opposite to each other and separated by a N atom from an intermediate layer. This defect features a low-energy phonon replica shifted just $\sim 62$~meV, see Fig.~\ref{fig:Phonon}A. In this defect, the C atoms are slightly displaced out of the plane (emulating a C--C repulsion) in the ground state, and the phonon associated with such distortion is responsible for the 62~meV phonon replica. A similar scenario occurs in C$_B$C$_B$--$(1,1)$. Fig.~\ref{fig:Phonon}B, C, shows the phonon with the most relevant partial Huang-Rhys factor $S_k$, for two different dimers, C$_B$C$_B$--$(\sqrt{3},2)$ and C$_B$C$_B$--$(0,2)$, while in the former the electronic excitation induces are rearrangement restricted to the defect plane, in the later the dominant phonon modes are out of plane. 

To the best of our knowledge, there are a couple of reports of defects in hBN with a similar low-energy phonon replica, the so-called yellow emitters \cite{Kumar2023}, but these emitters lack spin activity. Although we can discard C$_B$C$_B$--$(0,2)$ and  C$_B$C$_B$--$(1,1)$ as yellow emitters, it is likely that the yellow emitters trigger an out-of-plane distortion, explaining their low-energy phonon replicas. The other report of SPEs with a similar PL was obtained in a hBN flake after a H$_2$ plasma treatment \cite{chen2021generation}.

\subsection{Weakly coupled spin pairs}

Several reports of SPEs in hBN exhibit optically detected magnetic resonance (ODMR) with very small zero-field splittings (ZFS) in the MHz range \cite{Stern2022, Yang2023Laser, Guo2023coherent, chejanovsky2021}. Such small ZFS values can arise either from weakly coupled, physically distinct $S=1/2$ defects \cite{Scholten2024} or from individual defects possessing a triplet ground state with an intrinsically small ZFS. Our calculated C$_X$C$_X$ defects are candidates for both scenarios: when the two carbon atoms are separated by 0.4-0.7 nm, the ZFS within their triplet state is expected to be small, primarily due to the $r^{-3}$  decay of the magnetic dipole-dipole interaction. It is important to note that a small ZFS within the triplet manifold is not incompatible with a large exchange energy ($J$) that separates the singlet and triplet states, a situation relevant for our C$_X$C$_X$ defects at shorter C-C distances (see Fig.~\ref{fig:ST}). At larger separations, as the defects become spin degenerate.

At this point, it is informative to estimate the ZFS of the C$_X$C$_X$ defects where the constituent carbon atoms are separated by distances $r$ in the range of 0.4-1.0 nm; the magnetic dipole-dipole interaction is expected to be a primary contributor. Using the point-dipole approximation ($D_{dd}\approx C/r^{3}$, with $C\approx78$ MHz~nm$^3$, see Section V from the supplementary material), we estimate the ZFS parameter $D$ to range from approximately 78 MHz (for $r=1.0$ nm) up to 1.22 GHz (for $r=0.4$ nm). For a specific configuration like C$_B$C$_B$--$(0,2)$ with an inter-carbon distance of 0.67 nm, this estimation yields a $D$ value of approximately 260 MHz. These values, particularly at the larger end of the separation range, are substantially smaller than the multi-GHz ZFS typically observed for compact spin-1 defects like V$_B^-$ in hBN \cite{Gottscholl2021b}, and fall within a range that could be consistent with ODMR experiments reporting MHz-scale features.

A general model by Robertson \textit{et al.} \cite{robertson2024universal} proposes that some of the experimentally observed ODMR signals arise from pairs of weakly coupled $S=1/2$ defects. While our C$_N$C$_B$--$(d,h)$ defects share some characteristics with the systems considered by this model, such as C-C distances around $\sim 1$ nm, their established singlet ground state (due to charge transfer) makes them unlikely candidates to be these $S=1/2$ pairs themselves.

This distinction motivates exploring other defect types. Here, we propose that our C$_X$C$_X$ homonuclear defect pairs, provide a compelling alternative origin for such ODMR signatures. As discussed, when these C$_X$C$_X$ defects are sufficiently far apart, the exchange energy ($J$) diminishes, and the system behaves as two weakly interacting $S=1/2$ centers. The resulting state would exhibit a very small ZFS, as argued above. Such a system could indeed produce ODMR signals that appear to arise from `effectively $S=1/2$ defects,' consistent with the experimental findings of Scholten \textit{et al.} \cite{Scholten2024}, who observed this characteristic behavior through Rabi oscillation measurements. Thus, the C$_X$C$_X$ pairs at in hBN multilayers naturally reconcile the picture of weakly interacting $S=1/2$ spins with an underlying triplet state possessing a minimal ZFS.

\subsection{Suitability or usefulness as SPEs}

Overall, the studied defects present a mix of typical behaviors for in-plane defects and new features. Heteronuclear C$_B$C$_N$ defect pairs are predicted to behave as the typical bright SPEs usually found in hBN \cite{Tran2015}. They have a typical phonon sideband with high-energy replicas (160-190 meV), and their emission covers the entire visible range, also they usually do not produce a measurable ODMR signal \cite{cakan2024quantum}. All these characteristics suggest a large family of related defects, as proposed by different arrangements of substitutions C defects acting as donor-acceptor systems \cite{jara2021, Auburger21}. Our contribution shows that in such a family, C defects residing in different layers are also candidates for SPEs.

Instead, the homonuclear defects C$_B$C$_B$, C$_N$C$_N$ show some interesting features for SPEs, \textit{(i)} often they are spin-active at room temperature (\textit{i.e.}, a triplet) and \textit{(ii)} regardless of details, each series (C$_B$C$_B$ or C$_N$C$_N$) has a very similar ZPL, converging to an asymptotic limit. Both features are not present in C dimers within the same monolayer \cite{Pinilla} and are interesting for SPEs. On one hand, most of the already known spin-active defects in hBN are based on vacancies \cite{Liu2022}, which share many favorable and detrimental aspects with well-studied SPEs such as the NV-center in diamond (\textit{e.g.}, triplet ground state, ZFS in the GHz range, broad emission line). Instead, most defects spin-active C$_B$C$_B$, C$_N$C$_N$ defects would have some interesting properties as SPEs: a bright emission, a narrow emission line, and well-separated phonon replicas. But, their ZFS would be non-constant and range from tens to hundreds of MHz, which could make the integration with current technology difficult.

There are two particular defect configurations, C$_B$C$_B$--$(0,2)$ and  C$_B$C$_B$--$(1,1)$, with a phonon sideband different, to the best of our knowledge, from other substitutional C-based SPEs, characterized by prominent $\sim 60$ meV replicas from out-of-plane modes. If this distinct vibronic structure leads to an effectively broader overall emission profile where the ZPL is less dominant or more difficult to spectrally isolate, it could hinder applications requiring high photon indistinguishability and long coherence times, as these rely heavily on the properties of the ZPL photons.

Arguably, the most significant problem of SPEs in hBN is the difficulty of making monochromatic SPEs within the visible. So far, there are two irradiation protocols to achieve monodisperse emitters: the so-called blue \cite{Gale22} and the yellow \cite{Kumar2023} emitters. In this respect, very specific defect geometries -such as C$_B$C$_B$--$(0,2)$- are unlikely to be deterministically created. Instead, there are irradiation protocols aimed at producing $V_B$  defects \cite{carbone2025}. Such protocols could be adapted \cite{Ren2025} to produce pairs of $V_B$ within less than a nm. In the presence of carbon contamination \cite{mendelson2021}, these defects could turn into C$_B$C$_B$ dimers as those studied here, some of them with a triplet ground state. 

Are the pairs of C defects considered here SPEs at all? For strongly coupled C$_X$C$_X$ pairs forming a triplet state, or for C$_N$C$_B$ singlet pairs, the pair acts as a single quantum emitter. In the limit of weakly interacting C$_X$C$_X$ pairs, while the two $S=1/2$ centers are nearly independent, single-photon emission would still be expected if the optical cycle involves defined joint states of the pair. Definitive confirmation of single-photon emission characteristics relies on $g^{(2)}(\tau)$ measurements. As stated before, measurements of hBN defects of yet undetermined atomic structure reveal a very small ZFS tens of MHz or even smaller. These reports routinely show the dip $g^{(2)}(\tau)$ associated with single-photon emission. This is a strong hint that well-separated $S=1/2$ defects can behave as SPEs. Particularly, one article \cite{Stern2022} shows a defect in hBN with a ZFS of 50 MHz, together with a photoluminescence spectrum featuring the typical high-energy phonon replicas (180 meV) of C-based defects in hBN and a ZPL of $1.8$ eV. All these measurement points to C$_N$C$_N$ defects of the type studied here, perhaps with a larger C--C distance (1.1-1.2 nm). 

Even though we didn't calculate the emission rates of the defects studied here, it is relevant to ask if these weakly interacting $S=1/2$ pairs have an emission rate relevant for applications (or even detection). We can not give a definitive answer. However, the experiments mentioned above suggest a large enough emission rate, as the single defects provide a clear signal. For the donor-acceptor-like pairs, the charge transfer should imply a large dipole transition moment and a large emission rate. Experimentally, their ``brightness is the highest reported for a quantum emitter in the visible spectral range'' \cite{Tran2015}.

\section{\label{sec:conclusions}Conclusions}

Substitutional carbon defects are the most widely accepted explanation for the bright light emitters found in hBN. To the best of our knowledge, C-based defects have only been studied when in the same layer. 

In this article, we show, by means of DFT calculations, that neutral C defects lying in different layers are also viable candidates to be associated with single emitters. If they form donor-acceptor-like pairs, they behave similarly to when residing in the same hBN layer: they are spinless, have a ZPL energy of $\sim 1.9-2.4$ eV, and their photoluminescence spectra feature well-defined high-energy phonon replicas. 

However, if the two C atoms in different hBN layers are substituting the same species, N or B, they likely have a triplet ground state for closer distances $3.5-7.1$~\AA. It is separated by more than $\sim 0.5$ eV from the singlet state, making these defects spin-active above room temperature. Their ZPL is almost independent of the C-C distance (\textit{i.e.} nearly monochromatic), according to SCAN-based calculations, their emission is within the red. Calculations with the HSE06 functional shift the emission energy to the yellow-to-green range. Regardless of the actual value of the ZPL, the nearly monochromatic emission holds. Most of these kinds of defect dimers present the typical high-energy phonon replicas in their photoluminescence spectra.

As the distance of the mononuclear defects (C$_B$C$_B$ or C$_N$C$_N$) increases, they become practically spin degenerate, or `weakly interacting spin $1/2$ pairs' \cite{robertson2024universal}. In either case, a robust triplet or a spin degenerate state, the defects are well-separated and their zero field splitting should be within the MHz range, which is consistent with some optically detected magnetic resonance (ODMR) experiments \cite{Scholten2024,Stern2022, Yang2023Laser, Guo2023coherent, chejanovsky2021}. 

When two C$_B$ atoms are on top of each other but separated by a hBN interlayer or lying as close as possible in adjacent layers -C$_B$C$_B$--$(0,2)$ and C$_B$C$_B$-$(1,1)$- their photoluminescence spectrum has a markedly different phonon sideband, with well-defined phonon replicas at $\sim 60$ meV. They correspond to out-of-plane rearrangement upon excitations. We are unaware of any measurement of such a spectrum for spin-active defects, but the so-called (spinless) yellow emitters \cite{Kumar2023} have similar phonon replicas and could also correspond to out-of-plane distortions.

\section{Competing Interests}
The Authors declare no Competing Financial or Non-Financial Interests

\section{Data Availability}
Input files for the most relevant calculations are provided as supplementary information. The remaining data is available from the corresponding author upon reasonable request.

\section{Author Contributions}
I.C. and F.M. ran all the calculations. F.M. conceived the idea. C.C., A.E., and I.C. explained the stability of the triplet state. All authors discussed the results and contributed to the manuscript's writing.

\begin{acknowledgments}
This work was partially supported by Fondecyt Grants No. 1191353, 1220715, 1220366, 3240387, and 1221512, by the Center for the Development of Nanoscience and Nanotechnology CEDENNA AFB180001, and from Conicyt PIA/Anillo ACT192023. This research was partially supported by the supercomputing infrastructure of the NLHPC (ECM-02).
\end{acknowledgments}

\bibliography{bib}% Produces the bibliography via BibTeX.

\end{document}